\documentclass[preprint,aps,superscriptaddress]{revtex4-1}

\usepackage{amsmath}
\usepackage{amssymb}
\usepackage{graphicx}
\usepackage{dcolumn}
\usepackage{bm}
\usepackage{color}

\begin{document}
\title{Diffusion coefficient of an inclusion in a liquid membrane 
supported by a solvent of arbitrary thickness}
\author{Kazuhiko Seki}
\affiliation{
National Institute of Advanced Industrial Science and Technology (AIST)\\
AIST Tsukuba Central 5, Higashi 1-1-1, Tsukuba, Ibaraki, Japan, 305-8565
}
\author{Sanoop Ramachandran}
\affiliation{
Department of Chemistry, 
Graduate School of Science and Engineering, 
Tokyo Metropolitan University, 
Tokyo 192-0397, Japan 
}
%
\author{Shigeyuki Komura
}
%
\affiliation{
Department of Chemistry, 
Graduate School of Science and Engineering, 
Tokyo Metropolitan University, 
Tokyo 192-0397, Japan 
}

\date{05/13 Received: date / Revised version: date}
%
\begin{abstract}
The diffusion coefficient of an inclusion in a liquid membrane 
is investigated by taking into account the interaction between membranes and bulk 
solvents of arbitrary thickness. 
As illustrative examples, the diffusion coefficients of two types of 
inclusions - a circular domain composed of fluid with the same viscosity as the 
host membrane and that of a polymer chain embedded in the membrane are studied.
The diffusion coefficients are expressed  in terms of the hydrodynamic 
screening lengths which vary according to the solvent thickness. 
When the membrane fluid is dragged by the solvent of finite thickness,
via stick boundary conditions, multiple hydrodynamic screening lengths together
with the weight factors to the diffusion coefficients are obtained from 
the characteristic equation. 
The condition for which the diffusion coefficients can be approximated by 
the expression including only a single hydrodynamic screening length 
are also shown. 
\end{abstract}
\maketitle

\section{Introduction}
\label{introduction}

Recent advances in experimental techniques have made the direct 
observation of the Brownian motion of $\mu$m sized objects in membranes using 
microscopy and imaging a routine process~\cite{Tanaka-07,yanagisawa-07,cicuta-07,kaizuka-04,reitz-01,gambin-06,Aliaskarisohi}.
As a result, diffusion coefficients can be measured accurately and 
it is now possible to address the issue of the differences between 
Brownian motion of macromolecules embedded in membranes in various environments. 

Vesicles with sizes of the order of $10$ $\mu$m are frequently used
in experiments while the typical distance of a supported membrane from the substrate 
is of the order of $20$ \AA~\cite{Tanaka-07,yanagisawa-07}.
Obviously, in both these general cases the coupling of membrane with 
its environment are very different and hence it will influence 
the Brownian motion of inclusions. 
In this paper, we investigate the influence of solvent environments on 
diffusion of an inclusion embedded in a membrane. 
In the biological context, there are many examples of membranes 
coming in contact with a solvent of various depth such as in tissues.

Biological membranes can be regarded as two-dimensional (2D) viscous fluids. 
An important feature of membranes as a transport media is that they are 
not purely isolated~\cite{izuyama-88,suzuki-89,evans-88}.
Liquid membranes are coupled to surrounding solvents by interaction of polar 
head groups of lipid molecules with solvents; they form quasi-2D systems 
coupled to three-dimensional (3D) solvents. 
The coupling to the surrounding environments induces the momentum exchange between 
the membrane and the solvents. 
The influence of the momentum exchange on the Brownian dynamics has been 
theoretically investigated by introducing a phenomenological coupling constant or 
simplifying the solvents 
flow~\cite{saffman-75,saffman-76,izuyama-88,suzuki-89,evans-88,hughes-81,seki-93,sanoop-drag-10,Petrov-08,komura-95}.
These studies have also been extended to investigate the concentration 
fluctuations~\cite{seki-07,Tserkovbyak-06,inaura-08}.

Despite the large number of studies, the Brownian motion of an object in liquid membranes has not yet been fully understood.  
In a hydrodynamic description, 2D flow in a bilayer membrane can be regarded 
as viscous and the interaction between liquid membranes and surrounding solvents 
can be taken into account by the stick boundary condition between them. 
Diffusion coefficients of macroscopic inclusions embedded in membranes were  
analytically investigated for a planar membrane surrounded by solvent layers 
of infinite \cite{saffman-75,saffman-76,hughes-81,Petrov-08} 
or very small thickness \cite{izuyama-88,suzuki-89,evans-88,seki-93,sanoop-drag-10,komura-95}. 
These studies revealed that the hydrodynamic flow in a membrane is screened 
by the solvent drag force and is characterized by a hydrodynamic screening length. 
When a planar membrane is surrounded by infinite thickness of solvent, 
it is called the Saffman and Delbr\"uck (SD) hydrodynamic screening length, $\nu^{-1}$, and is 
given by the ratio between the 2D membrane viscosity $\eta$ and the 
3D solvent viscosity $\eta_s$, $\nu^{-1}=\eta/\eta_s$~\cite{saffman-75,saffman-76}. 
(As we shall see below, the dimension of the 2D membrane viscosity is that of 
3D solvent viscosity times a length.) 
In the opposite limit of a thin solvent layer of the thickness $h$, 
Evans and Sackmann (ES) hydrodynamic screening length 
given by $\sqrt{h/\nu}$ is appropriate~\cite{evans-88}.  
In both limits, the diffusion coefficients depend logarithmically on the size 
of the inclusions as long as the size is smaller than the hydrodynamic 
screening length. 
On the other hand, the diffusion coefficients depend on the size of the inclusions 
very differently when the size of the inclusions exceeds the 
hydrodynamic screening length. 
These studies naturally lead to the interest in the hydrodynamic screening length 
and its influence on the diffusion coefficients when the solvent layer has a 
finite thickness.

The solvent flow can be varied by changing the solvent thickness. 
The flow of solvents influences the membrane flow through the stick boundary 
condition imposed between the membrane and the solvents. 
As a result, the diffusion coefficients depend on the solvent thickness. 
The influence of the finite solvent thickness has been recently studied for diffusion of a disk~\cite{stone-98}, 
concentration fluctuations~\cite{Haataja-09,inaura-08,sanoop-11}, 
correlated diffusion~\cite{oppenheimer-09,diamant-09b,oppenheimer-10,sanoop-bulk-10}, 
 and polymer diffusion in a 
membrane~\cite{Ramachandran-20}.
Diffusion coefficients of other types of inclusions on 
membranes~\cite{levine-04,levine-04b,muthukumar-85,Naji-07,komura-95} or on Langmuir 
monolayers~\cite{fischer-04} have also been theoretically calculated.
However, the investigation on the influence of finite thickness of solvent 
was limited to numerical evaluation of the diffusion coefficients, where 
the dependence of the hydrodynamic screening length on the solvent thickness 
was not completely elucidated \cite{oppenheimer-09,diamant-09b,Ramachandran-20,inaura-08,Haataja-09,sanoop-bulk-10,stone-98,sanoop-11,oppenheimer-10}.  
In this paper, the relation between the diffusion coefficients and the hydrodynamic 
screening lengths are throughly investigated for an arbitrary thickness of the 
solvent layers on the basis of the analytical expression on the hydrodynamic 
screening lengths.

The relation between the diffusion coefficients and the hydrodynamic screening 
lengths can be shown in a straight-forward manner for a polymer embedded 
in a membrane by the Zimm model, where the equilibrium average of the 
hydrodynamic interactions is performed in 2D~\cite{komura-95,muthukumar-85,Ramachandran-20}. 
The multiple hydrodynamic screening lengths are then found for the 
finite solvent thickness.
The diffusion coefficients are expressed by the weighted sum; 
each term in the sum is a product of the weight factor and the function of the 
dimensionless size of the polymer normalized by each hydrodynamic screening length. 
On the basis of the analytical expression, the condition that the diffusion 
coefficient is approximately represented solely by the ES hydrodynamic screening 
length can be discussed in detail. 
We show that the diffusion coefficient cannot be approximated by the ES hydrodynamic 
screening length when both $\nu^{-1}=\eta/\eta_s$ and the size of the macromolecule 
are smaller than the solvent thickness.

Essentially the same relation between the diffusion coefficients and the hydrodynamic 
screening lengths is obtained for diffusion of a circular liquid domain with the 
same viscosity as that of the host membrane. 
The diffusion coefficient of a circular liquid domain embedded in a membrane has been 
studied in relation to recently proposed raft model, where rafts are formed by 
sphingomyelin and cholesterol rich liquid 
domains~\cite{simons-97,brown-98,klingler-93,oradd-05,cicuta-07,yanagisawa-07,kenworthy-04,Aliaskarisohi}. 
It is believed that rafts undergo lateral Brownian motion within a bilayer membrane 
and act as platforms for protein association and signaling \cite{brown-98}. 
Previously, the diffusion coefficient of a circular liquid domain of arbitrary size 
was derived in the limit of infinite depth of solvent layer or the limit of small 
depth of solvent 
layer~\cite{DeKoker,sanoop-drag-10,Fujitani,Aliaskarisohi}. 
In this paper, the results are generalized for the arbitrary thickness of solvent 
layers.
The diffusion coefficient is obtained as a simple integral which can be expressed 
again as the sum of the terms given by functions of the same hydrodynamic screening 
lengths multiplied by the same weight factors as those for the polymer diffusion 
coefficients.

In Sec.~\ref{Hydrodynamics}, the membrane hydrodynamics is reviewed. 
The diffusion coefficient of a polymer embedded in a membrane is obtained in 
Sec.~\ref{polymer}. 
In Sec.~\ref{screening}, the relation between hydrodynamic screening length and 
the solvent thickness is discussed. 
The diffusion coefficient of a liquid domain in a membrane is obtained in 
Sec.~\ref{Calculation}. 
Finally, the last section is devoted to conclusions.

\section{Hydrodynamic flow in a membrane and solvent}
\label{Hydrodynamics}

\begin{figure}
\centerline{
\includegraphics[width=0.5\columnwidth]{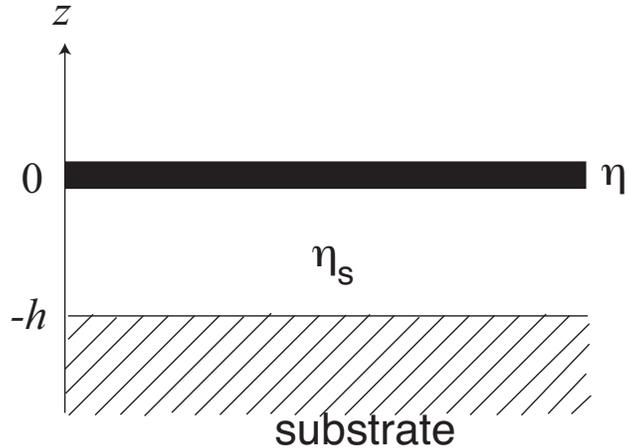}
}
\caption{Schematic picture showing a planar liquid membrane with 
2D viscosity $\eta$ located at $z=0$.
It is supported on a solvent with 3D viscosity $\eta_{\rm s}$.
A substrate is located at $z=- h$ bounding the 
solvent.} 
\label{fig:membwall}
\end{figure}

As shown in Fig.~\ref{fig:membwall}, we consider the situation where the 
liquid membrane is supported by a bulk solvent on the solid substrate.
 The situation where the membrane is also supported by a solvent from above 
will be considered in Sec. VI. 
We denote the 2D flow in the membrane by ${\bf v}({\bf r})$ where  
${\bf r}=(x,y)$ represents a position within the plane of the 
membrane.
The membrane is regarded to be incompressible,   
\begin{equation}
{\bf \nabla} \cdot {\bf v} ({\bf r})= 0.  
\label{eqn:2Dcompress}
\end{equation}
Here $\nabla$ is a differential operator in the 2D  Euclidean space. 
The viscous flow in the membrane can be expressed by the Stokes equation in 2D, 
\begin{equation}
\eta \nabla^2 {\bf v}({\bf r}) - \nabla p({\bf r}) + {\bf f}_{\rm s} ({\bf r})=0,
\label{eqn:2Dstokes}
\end{equation}
where $\eta$ is the 2D membrane viscosity,
$p({\bf r})$ the in-plane pressure, and 
${\bf f}_{\rm s}({\bf r})$ the in-plane force exerted on the membrane from the 
solvent.
The last quantity can be obtained when the solvent fluid velocities are determined. 
The stress tensor of the liquid membrane is given by  
\begin{equation}
\sigma_{\alpha \beta} ({\bf r})= 
- p({\bf r}) \delta_{\alpha \beta} 
+ \eta [\partial_\alpha v_\beta({\bf r})
+ \partial_\beta v_\alpha({\bf r})], 
\label{eqn:stress2D}
\end{equation}
where $\delta_{\alpha \beta}$ is the Kronecker delta,  
and $\alpha$, $\beta$ are $x$, $y$.
Then Eq.~(\ref{eqn:2Dstokes}) can be represented in terms of the stress 
tensor as, 
\begin{align}
{\rm div} \, {\bm \sigma} + {\bm f}_{\rm s} ({\bm r})  =0 ,  
\label{eqn:"Dstokes1}
\end{align} 
where $({\rm div} \, {\bm \sigma})_\alpha = \sum_\beta \partial  \sigma_{\alpha \beta} /\partial x_\beta$.

As shown in Fig.~\ref{fig:membwall}, the membrane is located in the plane at $z=0$.
The solvent velocities ${\bf v}^{(3)}({\bf r},z)$, satisfy the incompressibility 
condition
\begin{equation}
\tilde{{\bf \nabla}} \cdot {\bf v}^{(3)} ({\bf r},z) = 0,
\label{eqn:3Dcompress}
\end{equation}
where $\tilde \nabla$ represents a differential operator in the 3D Euclidean space. 
We denote the 3D viscosity of the solvent as $\eta_{\rm s}$, and the solvent flow 
also obeys the 3D Stokes equation, 
\begin{equation}
\eta_{\rm s} \tilde{{\bf \nabla}}^2 {\bf v}^{(3)} ({\bf r},z)- 
\tilde{{\bf \nabla}} p^{(3)} ({\bf r},z)= 0, 
\label{eqn:3Dstokes}
\end{equation}
where $p^{(3)}({\bf r},z)$ represents the pressure of the solvent. 
The solvent is supported on the substrate which is located at $z=- h$. 
The no-slip boundary condition is imposed at $z=- h$ as well as between the 
membrane flow and the solvent flow. 
Through this boundary condition, the surrounding solvent exerts a drag force on 
the liquid membrane.

The drag force in Eq.~(\ref{eqn:2Dstokes}) can be expressed as  
${\bf f}_{\rm s} = -({\bf I} - \hat{{\bf e}}_z \hat{{\bf e}}_z) \cdot
{\bf \sigma}^{(3)}({\bf r},0) \cdot \hat{{\bf e}}_z$, 
where $\hat{{\bf e}}_z$ is the unit vector along the $z$-axis. 
The tensorial component of ${\bf I}$ is given by $\delta_{ij}$, and 
${\bf I}- \hat{{\bf e}}_z \hat{{\bf e}}_z$ denotes the projection to 
the in-plane space. 
The stress tensor of solvent is given by 
\begin{equation}
\sigma_{i j}^{(3)} ({\bf r},z)= 
- p^{(3)} ({\bf r},z)\delta_{i j} 
+ \eta_{\rm s} [\tilde{\partial}_i v_j^{(3)} ({\bf r},z)
+ \tilde{\partial}_j v_i^{(3)}({\bf r},z)], 
\label{eqn:stress}
\end{equation}
where $i$, $j$ denote $x$, $y$, $z$.

Using the stick boundary conditions at $z=0$ and $z=- h$, we solve the 
hydrodynamic equations from Eq.~(\ref{eqn:3Dcompress}) to Eq.~(\ref{eqn:3Dstokes})  
to obtain ${\bf f}_{\rm s}$. 
In the Fourier space, ${\bf f}_{\rm s}$ is calculated to 
be~\cite{inaura-08,fischer-04,lubensky-96}
\begin{equation}
{\bf f}_{\rm s}[{\bf k}] = 
- \eta_{\rm s} k \coth(kh) {\bf v}[{\bf k}], 
\label{eq:fs}
\end{equation}
where ${\bf k}=(k_x,k_y)$ and $k = \vert {\bf k} \vert$.
The real space velocity field of the membrane flow ${\bf v}({\bf r})$ can be 
expressed as 
\begin{equation}
{\bf v}({\bf r}) = 
\int \frac{{\rm d}^2 k}{(2\pi)^2} \,
{\bf v}[{\bf k}]
\exp ( i {\bf k}\cdot {\bf r}). 
\label{eqn:FT}
\end{equation}
The Fourier space mobility tensor ${\bf G}[{\bf k}]$ associated with the velocity 
field is given by~\cite{fischer-04,lubensky-96,inaura-08}
\begin{equation}
G_{\alpha \beta}[{\bf k}] =  \frac{1}{\eta [ k^2 + \nu k \coth(kh)]} 
\left( \delta_{\alpha\beta} - \frac{k_\alpha k_\beta}{k^2} \right).
\label{eqn:cothoseen}
\end{equation}

In order to calculate diffusion coefficients, the mobility tensor in Fourier space 
should be transformed into real space. 
Previously, the inverse Fourier transform of the mobility tensor was analytically 
performed only in the limits of infinite or zero thicknesses of a solvent layer. 
In the next section, the inverse Fourier transformation of the mobility tensor is 
analytically performed for an arbitrary thickness of a solvent.

\section{Diffusion coefficient of a 2-dimensional polymer chain}
\label{polymer}

As an illustrative example for the influence of finite thickness of solvent 
on the diffusion coefficient of a macromolecule embedded in a 2D planar membrane, 
we consider the diffusion of a polymer chain confined in the 
membrane~\cite{muthukumar-85,komura-95,Ramachandran-20}.
Previously, the influence of the solvent on diffusion coefficients is 
analytically investigated only in the limits of very thin or infinite 
thicknesses of solvent layers. 
In these works, the hydrodynamic screening length is a key quantity 
in characterizing the screening of the flow of membrane by the presence 
of solvent layers. 
The influence of finite thickness of solvent was investigated by numerically 
evaluating the inverse Fourier transform of the mobility tensor, where 
the hydrodynamic screening length was not even defined. 
In this section, the hydrodynamic screening lengths are obtained from an 
analytical equation for arbitrary thickness of solvent layer.

The conformation of a 2D polymer chain embedded in a 2D membrane is 
represented by $N$ beads with position vectors, 
$\{ {\bf R}_n \} = ({\bf R}_1, \ldots, {\bf R}_N)$,  
under the potential energy, 
\begin{equation}
U= \frac{ k_{\rm B}T}{b^2} 
\sum_{n=2}^N ({\bf R}_n - {\bf R}_{n-1})^2,
\label{eqn:polypot}
\end{equation}
where $b$ is the Kuhn length \cite{doi-edwards}.
The mobility tensor associated with the beads is 
given by the inverse Fourier transform of Eq.~(\ref{eqn:cothoseen}) as 
\begin{equation}
G_{\alpha\beta}\left(  {\bf R}_n- {\bf R}_m \right)= 
\int \frac{{\rm d}^2 k}{(2\pi)^2} \,
G_{\alpha \beta}[{\bf k}] \exp 
\left[ i {\bf k} \cdot\left( {\bf R}_n- {\bf R}_m \right) \right].  
\label{polymerG}
\end{equation}
Within the pre-averaging approximation \cite{doi-edwards}, the polymer diffusion 
coefficient is expressed as
\begin{equation}
D_{\rm poly} = k_{\rm B}T \int_0^N  \frac{{\rm d}n}{N} \int_0^N  \frac{{\rm d}m}{N} 
\, g(n-m), 
\label{eqn:genD}
\end{equation}
where 
$g(n-m)$ is the isotropic component of mobility tensor 
$G_{\alpha\beta} \left( {\bf R}_n- {\bf R}_m \right)$~\cite{Ramachandran-20}.
By using Eq.~(\ref{polymerG}), two analytical expressions for the diffusion coefficients have been derived from 
Eq. (\ref{eqn:genD}) in the limits of very thin or infinite thickness of solvent layers
~\cite{komura-95,Ramachandran-20}. 
Here, we investigate the diffusion coefficient by keeping the finite depth of the solvent layer 
without taking the limits. 
By expanding $1/[k + \nu \coth(kh)]$ in partial fractions, 
we note the general relation~\cite{John-50} 
\begin{equation}
\int_0^\infty {\rm d}k \, \frac{f(k)}{k + \nu \coth(kh)} = 
\sum_{j=1}^\infty C_j \int_0^\infty {\rm d}k \, \frac{kf(k)}{k^2 + \kappa_j^2}, 
\label{eqn:integration}
\end{equation}
where $f(k)$ is an arbitrary function, $\kappa_j$ and $C_j$ will be later given 
by Eqs.~(\ref{eq:kappaj}) and (\ref{eq:Cj}), respectively.
By introducing Eq.~(\ref{eqn:integration}) into Eqs.~(\ref{polymerG}) and 
(\ref{eqn:genD}), we obtain, 
\begin{align}
g(n-m) &=- \sum_{j=1}^\infty C_j \frac{1}{8\pi\eta}
\exp\left(  \frac{1}{4} b^2 \kappa_j^2 \vert n-m \vert \right)
\nonumber\\ &\quad\quad\quad\quad
\times{\rm Ei}\left(- \frac{1}{4}b^2 \kappa_j ^2 \vert n-m \vert \right),  
\label{eqn:esmobility}
\end{align}
where ${\rm Ei}(-z)$ is the exponential integral~\cite{abram-stegun}.

In the real space, the mobility tensor is expressed in terms of an
infinite number of characteristic lengths, $\kappa_j^{-1}$, where $\kappa_j$ 
is determined by the following characteristic equation 
\begin{equation}
\cot (\kappa_j h) = \frac{\kappa_j}{\nu}.
\label{eq:kappaj}
\end{equation}
All the roots of the equation are given by 
$\kappa = \pm \kappa_j$ with $j=1,2, \cdots$.
 The characteristic lengths relative to $h$, $1/(\kappa_j h)$, 
depend on $\nu h$ given by the viscosity ratio $\nu=\eta_s/\eta$, 
and represent the screening of hydrodynamic flow in 2D membrane due to the 
presence of the solvent. 
The contribution of each screening length is weighted by the factor 
\begin{equation}
C_j = \frac{2\nu}{h \kappa_j^2+h \nu^2 + \nu}.  
\label{eq:Cj}
\end{equation}

Using Eq.~(\ref{eqn:esmobility}), the diffusion coefficient is obtained as  
\begin{align}
D_{\rm poly}&=\sum_{j=1}^\infty C_j 
\frac{k_{\rm B}T}{4 \pi \eta}
\frac{1}{\epsilon_j^4}
\Big[
(1+\epsilon_j^2)(2 \ln \epsilon_j + \gamma)
\nonumber\\&\quad\quad\quad\quad\quad\quad
- \epsilon_j^2 
-\exp(\epsilon_j^2){\rm Ei}(-\epsilon_j^2)
\Big],
\label{eqn:Des}
\end{align}
where $\gamma=0.5772 \cdots$ is Euler's constant.
In the above, we have defined the dimensionless polymer size as 
$\epsilon_j \equiv \sqrt{N}b \kappa_j/2 = R_{\rm g} \kappa_j$, and 
$R_{\rm g} = \sqrt{N}b/2$ is the radius of gyration for the 2D Gaussian 
polymer chain.

The limiting expression for $\epsilon_1 \ll 1$ is 
\begin{equation}
D_{\rm poly} \approx C_1
\frac{k_{\rm B}T}{4 \pi \eta}
\left[-\ln \epsilon_1- \frac{\gamma}{2} 
+ \frac{3}{4} \right]. 
\label{eqn:Dessmalle}
\end{equation}
As will be discussed in the next section, the above expression is close to the 
exact result under the additional condition of $\nu h <1$ which is 
needed to replace the sum in Eq.~(\ref{eqn:Des}) with the term related to 
$\epsilon_1$. 
When $\epsilon_1 \gg 1$, Eq.~(\ref{eqn:Des}) reduces to
\begin{equation}
D_{\rm poly} \approx C_1
\frac{k_{\rm B}T}{4 \pi \eta}
\frac{1}{\epsilon_1^2}
\Big[2 \ln\epsilon_1+\gamma-1\Big].
\label{eqn:Deslargee}
\end{equation}
This expression holds regardless of the value of $h$ as long as it is finite. 
The sum in Eq.~(\ref{eqn:Des}) can be represented by the single dominant term 
as long as $\epsilon_1 \gg 1$.
However, the additional condition of $\nu h <1$ is required when 
$\epsilon_1 \ll 1$, about which we shall discuss in the next section.

\section{Hydrodynamic screening length vs. solvent thickness}
\label{screening}

\begin{figure}
\centerline{
\includegraphics[width=0.5\columnwidth]{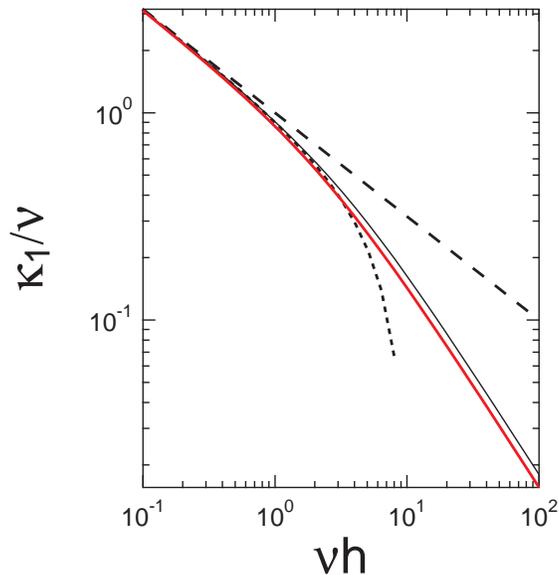}
}
\caption{ (Color online) The smallest positive values for the inverse of the characteristic lengths against 
the solvent thickness. Both quantities are normalized by $\nu=\eta_{\rm s}/\eta$. 
The red thick line represents  
 the smallest positive root of the characteristic equation, 
Eq. (\ref{eq:kappaj}), calculated numerically.
The long dashed line represents $\kappa/ \nu=1/ \sqrt{\nu h}$. The short dashed line 
represents the result of Eq. (\ref{eq:kappa11}). 
The thin solid line represents the result of Eq. (\ref{eq:kappa1}). 
} 
\label{fig:kappa}
\end{figure}

If the approximated diffusion coefficient obtained by taking into account 
only the smallest positive value of $\kappa_j$ (denoted by $\kappa_1$)  
reproduce the exact results, then $\kappa_1^{-1}$ can be regarded as the 
effective hydrodynamic screening length.

First, we consider the value of $\kappa_1$ which is the inverse of the effective 
hydrodynamic screening length as long as the higher order ($j \ge 2$) terms can 
be ignored. 
We first note the series expansion,  
\begin{equation}
x \cot x = 1 + 2 \sum_{n=1}^\infty \frac{x^2}{x^2-n^2 \pi^2}.
\end{equation}
Since the lowest order term can be estimated as 
$1+2 x^2/(x^2-\pi^2) \approx x^2/(\nu h)$, the approximate expression for 
$\kappa_1$ turns out to be    
\begin{equation}
\kappa_1 \approx  
\left(
\frac{3 \nu h + \pi^2 - \sqrt{(3 \nu h)^2+ 2 \nu h \pi^2 + \pi^4}}{2 h^2}
\right)^{1/2}. 
\label{eq:kappa1}
\end{equation}
In the limit of $\nu h/\pi^2<1$, $\kappa_1$ can be further approximated as 
\begin{equation}
\kappa_1 \approx 
\kappa\left(1- \frac{\nu h}{\pi^2} \right), 
\label{eq:kappa11}
\end{equation}
where $\kappa=\sqrt{\nu/h}$ is the inverse of the 
ES hydrodynamic screening 
length defined in the limit of $h \rightarrow 0$. 
In Fig.~\ref{fig:kappa}, the smallest positive values for the inverse of 
the characteristic lengths are presented against the solvent layer thickness, $h$. 
By increasing the solvent layer thickness $h$, the inverse of the hydrodynamic 
screening length rapidly decreases as shown in Fig.~\ref{fig:kappa}.

\begin{figure}
\centerline{
\includegraphics[width=0.5\columnwidth]{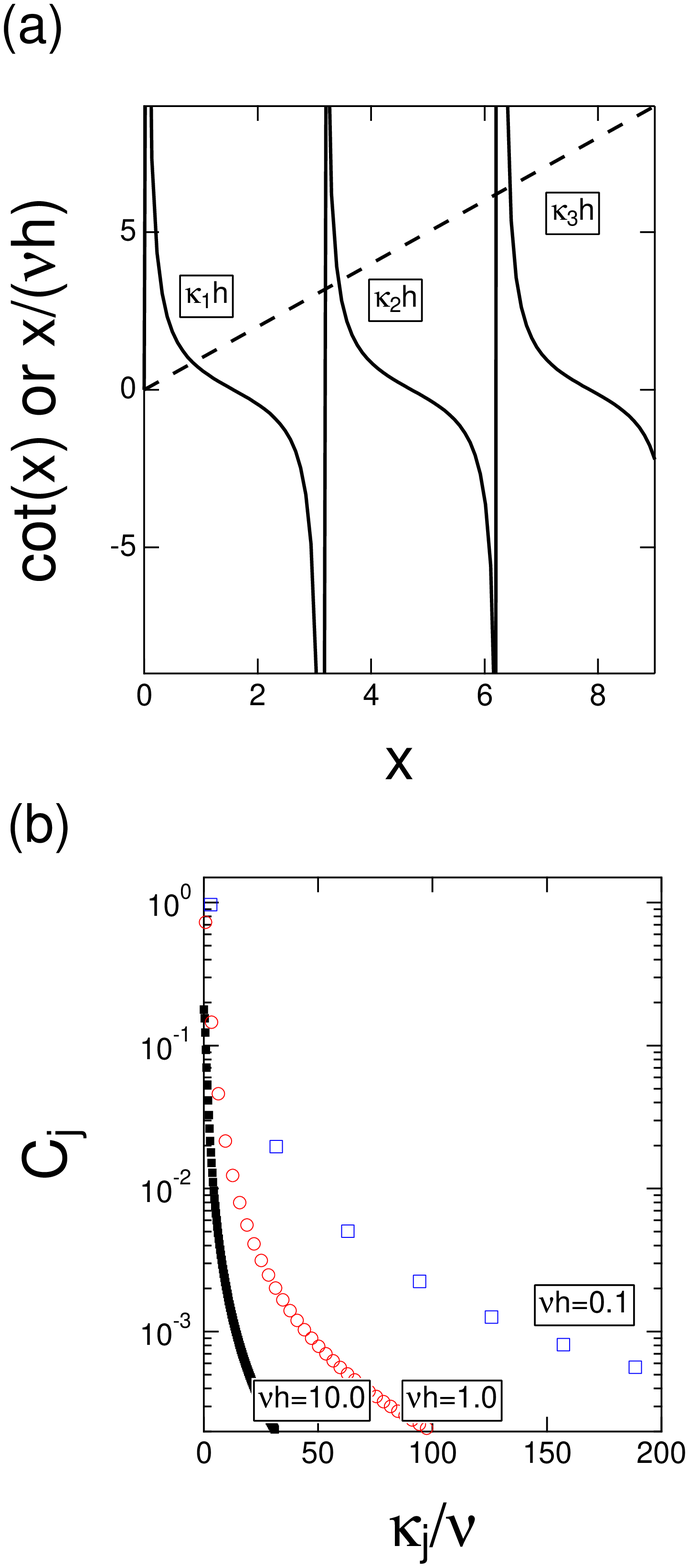}
}
\caption{ (Color online)}(a) The pictorial solution of the characteristic equation, Eq. (\ref{eq:kappaj});  
$\cot (x)$ against $x$ and $x/(\nu h)$ against $x$ for $\nu h=1$. The cross points of lines are 
$x_j=\kappa_j h$. The smallest positive value is $x_1$. 
$\kappa_1$ is obtained by  $\kappa_1=x_1/h$
(b) $C_j$ against $\kappa_j/\nu$. $C_j$ represents 
the weight associated with each hydrodynamic screening length, Eq.(\ref{eq:Cj}). 
Blue squares, red circles and dots represent $\nu h=0.1$, $\nu h=1.0$ and $\nu h=10.0$, respectively.  
\label{fig:dispersion}
\end{figure}

Next we consider the condition for which the diffusion coefficient can be 
characterized by a single hydrodynamic screening length as a good approximation 
for the exact expression including multiple hydrodynamic screening lengths 
associated with higher order $\kappa_j$. 
Judging from Eq.~(\ref{eq:kappaj}) and Fig.~\ref{fig:dispersion} (a), 
$\kappa_j$ takes discrete values which are almost equally separated. 
When $\kappa_1 R_{\rm g}$ is well separated from $\kappa_2 R_{\rm g}$ and the diffusion 
coefficient is given by the weighted sum of monotonically decreasing 
functions of $\kappa_j R_{\rm g}$ multiplied by the rapidly decreasing weights, 
the sum can be well represented by the term associated with $\kappa_1 R_{\rm g}$ alone.
Below, we show that $\kappa_1 R_{\rm g}$ is well separated from $\kappa_2 R_{\rm g}$ 
when $R_{\rm g}>h$ 
and the weights rapidly decay when $h< 1/\nu$. 

Since we have $\kappa_j \approx \kappa_1+\pi(j-1)/h$, the hydrodynamic 
screening lengths are separated by the factor $1/h$. 
Hence $\kappa_2 R_{\rm g}$ is well separated from $\kappa_1 R_{\rm g}$ when $R_{\rm g}/h >1$. 
It is convenient to define the cut-off size $R_{\rm g}^* = h$ over which the 
expression with $\kappa_1 R_{\rm g}$ could be very different from that with  
$\kappa_2 R_{\rm g}$.

In Fig.~\ref{fig:dispersion} (b), the weight factors $C_j$ are shown against $\kappa_j/\nu$. 
The weight factors $C_j$ in Eq.~(\ref{eq:Cj}) decrease with increasing 
$\kappa_j$.
The ratio of $C_2/C_1$ is an important factor in estimating whether the term 
related to $\kappa_1$ is dominant over other terms. 
Fig.~\ref{fig:dispersion} (b) shows that the difference between $C_1$ and $C_2$ 
increases by decreasing the thickness of solvent layer. 
Specifically, we have $\kappa_j \approx \kappa_1+\pi(j-1)/h$ and 
$C_2/C_1 \approx \kappa_1^2/\kappa_2^2 \approx \nu h /\pi^2$. 
Hence the diffusion coefficient can be approximated by the expression involving 
$\kappa_1$ alone when $\nu h /\pi^2 <1$. 
When the condition $\nu h <1$ is satisfied, 
$C_2/C_1 <1$ and the diffusion coefficients can be approximated by 
those obtained by Evans and Sackmann, where 
$C_1 \sim 1/(1+\nu h/2) \sim 1$. 
It is then convenient to define the critical thickness of solvent $h^* = 1/\nu$.
If the solvent depth exceeds $h^*$, the weight of $C_2$ is not much different 
from that of $C_1$. 
It should be noticed, however, that the precise estimation of the contribution 
from the higher modes requires the whole expression besides the weights. 

\begin{figure}%
\centerline{
\includegraphics[width=0.5\columnwidth]{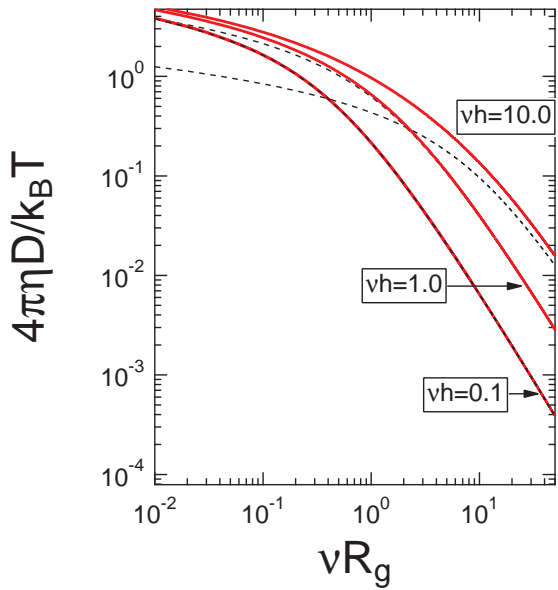}
}
\caption{ (Color online)The diffusion coefficients of polymer against size for various solvent thickness. 
The solid red lines indicate the complete solution obtained from Eq. (\ref{eqn:Des}).  
The short dashed black lines indicate the approximate solution by assuming a characteristic length scale for 
hydrodynamic screening given by $1/\kappa_1$. 
The short dashed line for $\nu h=0.1$ overlaps with the exact solution. 
} 
\label{fig:if_polymer}
\end{figure}

The expression of the diffusion coefficient depends on the kind of inclusions. 
As a representative example, we consider the diffusion coefficient of polymer   
to study conditions to use a single effective hydrodynamic screening length 
given by $1/\kappa_1$. 
In Fig.~\ref{fig:if_polymer}, we show the polymer diffusion coefficients 
against the size of the polymer $R_{\rm g} $ to study whether the polymer diffusion coefficients can 
be approximated by an expression without summation. 
When $\nu h \leq 1$, the polymer diffusion coefficients can be approximated 
by taking into account only $\kappa_1$ as shown in Fig.~\ref{fig:if_polymer}.
It is consistent with the fact that the weight $C_2$ is smaller than $C_1$ 
when $\nu h \leq 1$ since $C_2/C_1 \sim \nu h /\pi^2$. 
The situation corresponds to that considered by Evans and Sackmann. 

When $\nu h>1$, $C_1$ is close to $C_2$ and the functional form of the diffusion coefficient should be 
carefully examined.
When $\nu h>1$ and $R_{\rm g} >h$ holds, $\kappa_1 R_{\rm g}  >1$ is satisfied. 
Then the diffusion coefficient is well approximated by Eq.~(\ref{eqn:Deslargee}) 
showing $1/\left(\kappa_1 R_{\rm g} \right)^2$ dependence.   
Notice that $1/\left(\kappa_j R_{\rm g} \right)^2$ decays relatively fast by increasing $j$. 
When $\nu h>1$ and the size of the polymer $R_{\rm g} $ exceeds the solvent 
thickness $h$, the diffusion coefficient is approximated by the expression 
given in terms of $\kappa_1 R_{\rm g}$ alone.

When $\nu h>1$ and $R_{\rm g} <h$, on the other hand, a significant deviation 
is seen for the diffusion coefficients if the higher order terms are ignored, 
as can be seen from Fig.~\ref{fig:if_polymer}.   
This deviation originates from the fact that the weak logarithmic dependence 
on $\kappa_j R_{\rm g} $ and $\kappa_2 R_{\rm g} $ is not well separated from 
$\kappa_1 R_{\rm g} $ if $R_{\rm g} /h <1$. 
Also notice that $C_1$ is close to the other values of $C_j$ if $\nu h>1$. 
In such a situation, multiple hydrodynamic screening lengths should be taken 
into account.

To summarize, we have four length scales:  
the critical thickness of the solvent $h^*=\nu^{-1}$, 
the cut-off size of the polymer $R_{\rm g}^*=h$, 
the SD hydrodynamic screening length $\nu^{-1}$, and 
the ES hydrodynamic screening length 
$\kappa_1^{-1}$ (see Eq.~(\ref{eq:kappa1})). 
Although $h^*$ is identical to the SD hydrodynamic screening 
length $\nu^{-1}$, their physical meanings are different as explained below.

The diffusion coefficient can be approximately expressed by using either  
$\nu^{-1}$ or $\kappa_1^{-1}$. 
For the thin solvent layer, $h < 1/\nu$, the weights rapidly decrease with 
increasing $j$ and the diffusion coefficient is given by 
$\kappa_1^{-1}$ regardless of the macromolecule size. 
This is the limiting case considered by Evans and Sackmann. 
For thick solvent layers exceeding the critical thickness $h^*=1/\nu$, 
the diffusion coefficient can be approximated by the expression 
including only the single hydrodynamic screening length $\kappa_1^{-1}$, 
when the size of the macromolecule is larger than $R_{\rm g}^*=h$. 
In this case, we have $(\kappa_2 -\kappa_1) R_{\rm g} \sim R_{\rm g}/h >1$ and  
$\kappa_1 R_{\rm g}$ is well separated from $\kappa_2 R_{\rm g}$. 
When the solvent thickness exceeds $h^*=1/\nu$ and the size of macromolecules 
is smaller than $R_{\rm g}^*=h$, on the other hand, the diffusion coefficient shows 
weak logarithmic dependence on $\kappa_j R_{\rm g}$ and multiple hydrodynamic screening 
lengths should be taken into account.  
The diffusion coefficient is expressed by the hydrodynamic screening length 
$\nu^{-1}$ in the limit of $h \rightarrow \infty$.

\section{Diffusion coefficient of a circular domain}
\label{Calculation}

\begin{figure}
\centerline{
\includegraphics[width=0.5\columnwidth]{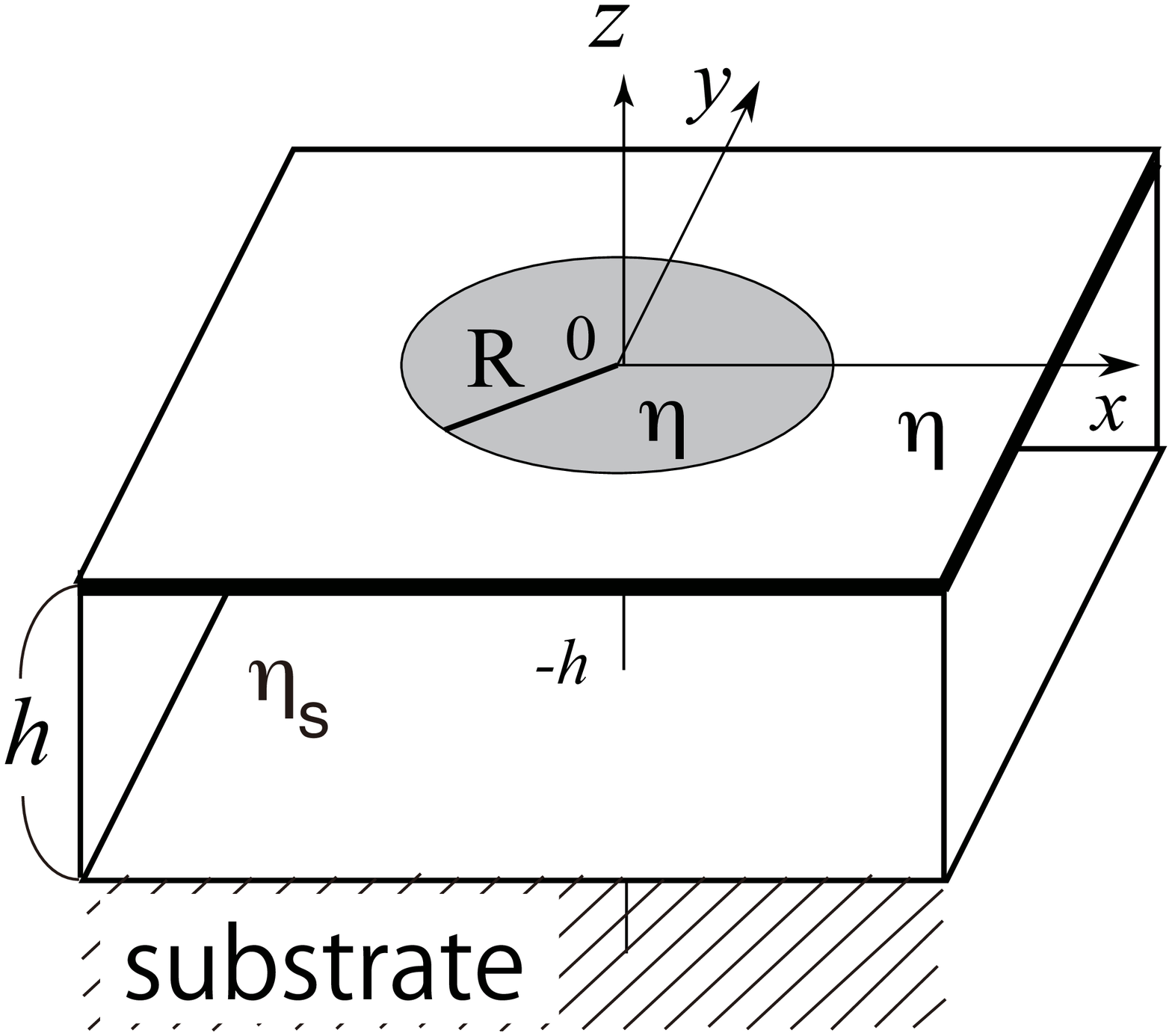}}
\caption{Schematic picture showing a liquid domain embedded in a planar 
liquid membrane located at $z=0$. 
Both a liquid domain and a membrane have the same  2D viscosity $\eta$. 
It is supported on a solvent with 3D viscosity $\eta_{\rm s}$.
A substrate is located at $z=- h$ bounding the solvent in the lower
region.} 
\label{fig:domain}
\end{figure}

In this section, we consider the diffusion coefficient of a circular liquid 
domain in a membrane (see Fig.~\ref{fig:domain}). 
Although the  characteristic equation, Eq.~(\ref{eq:kappaj}) associated with the 
solvent and the membrane flow should not be altered by changing the inclusion 
from polymers to liquid domains, the condition that the diffusion coefficient
can be approximated by truncating the infinite sum to a single expression 
depends on the size dependence of the diffusion coefficients. 
The size dependence can differ between polymers and liquid domains. 
For simplicity, we consider the case when the viscosity of the circular liquid 
domain is the same as that of the host membrane denoted by $\eta$. 
Previously, simple expressions for the diffusion coefficients were obtained 
for either infinite or very thin limits of the solvent 
layers~\cite{DeKoker,sanoop-drag-10}. 
Here we generalize the results to arbitrary thickness of solvent layers. 

We consider the situation for which the center of the circular object moves 
with the velocity ${\bf U}$, and its edge is assumed to keep circular shape 
without any deformation. 
The velocity field inside and outside the circular domain satisfy~\cite{stone-95}
\begin{equation}
\eta \nabla^2 {\bf v}({\bf r}) - \nabla p({\bf r}) 
+ {\bf f}_{\rm s}({\bf r}) + 
{\bf F}^{(\ell)}({\bf r}) \frac{\delta (r-R)}{2 \pi R} = 0, 
\label{eqn:2DstokesInd}
\end{equation}
and the incompressibility condition given by Eq.~(\ref{eqn:2Dcompress}) for 
all ${\bf r}$.
Here ${\bf f}_{\rm s}$ was defined before in Eq.~(\ref{eq:fs}), and   
${\bf F}^{(\ell)}$ is the force exerted at the periphery of the circle  
in the direction normal to the circular boundary~\cite{DeKoker}. 
If we take the origin of the coordinates at the center of the circular domain and   
choose the $x$-coordinate in the direction of ${\bf U}$, 
${\bf F}^{(\ell)}$ should vary according to the velocity ${\bf U}$ 
at the periphery of the circle.   
From the symmetry with respect to ${\bf U}$, ${\bf F}^{(\ell)}$ can 
be expressed as~\cite{DeKoker} 
\begin{equation}
{\bf F}^{(\ell)} ({\bf r})= F^{(\ell {\rm n})} \cos \theta \, {\bf n}, 
\label{Fs}
\end{equation}
where ${\bf n}={\bf r}/r$ is the outward normal unit vector at the surface of 
the circle of radius $R$, and $\theta$ is the angle between ${\bf U}$ and 
${\bf r}$.

Our task is to calculate the total force exerted on the circular domain in the steady 
state 
\begin{equation}
{\bm F} = \int_{r \leq R} {\rm d} {\bm r}\, {\bm f}_{\rm s} + 
\int_{r=R} {\rm d} \ell \, {\bm \sigma} \cdot {\bm n}, 
\label{eq:steadytforce}
\end{equation}
where ${\boldsymbol \sigma}$ is the stress tensor of the liquid membrane 
given by Eq.~(\ref{eqn:stress2D})  and $d \ell$ denotes the line integration. 
The first term represents the force exerted from the membrane flow
field, and the second term represents the direct friction force exerted 
from the solvent surrounding the membrane to the circular domain.

By using Gauss's theorem, we find 
\begin{align}
{\bm F}&= \int_{r=R} {\rm d} \ell \, {\bm \sigma} \cdot {\bm n} +
\int_{r \leq R} {\rm d} {\bm r} \, {\bm f}_{\rm s} \nonumber\\
& = \int_{r\leq R} {\rm d} {\bm r} \, \mbox{div} \, {\bm \sigma} +
\int_{r \leq R} {\rm d} {\bm r}\, {\bm f}_{\rm s} \nonumber \\
&= - \int_{r\leq R} {\rm d} {\bm r} \,
{\bm F}^{(\ell)} ({\bm r})\frac{\delta (r-R)}{2 \pi R} 
 \nonumber \\
&= -F^{(\ell {\rm n})} \hat{{\bm e}}_x \int_0^{2\pi} 
\frac{{\rm d} \theta}{2\pi} \, \cos^2 \theta 
=  -\frac{F^{(\ell {\rm n})}}{2} \hat{{\bm e}}_x, 
\label{eq:Gauss}
\end{align}
where Eq.~(\ref{eqn:"Dstokes1}) and Eq.~(\ref{eqn:2DstokesInd}) 
are used to obtain the third equality,  
and $\hat{{\bf e}}_\alpha$ denotes the unit vector along the 
$\alpha$-direction. 
Equation (\ref{eq:Gauss}) shows that it is sufficient to calculate 
$F^{(\ell {\rm n})}$ to obtain the total force exerted on the circular
object from the membrane flow field and the solvent.

The velocity field can be formally expressed as 
\begin{equation}
v_\alpha ({\bf r})= \int {\rm d} {\bf r}' \, 
G_{\alpha\beta} ({\bf r}-{\bf r}')
F_\beta^{(\ell)} ({\bf r}') \frac{\delta (r'-R)}{2 \pi R}. 
\label{vfield}
\end{equation}
In real space, the mobility tensor is expressed by the 
Fourier transform of Eq.~(\ref{eqn:cothoseen}) as
\begin{equation}
G_{\alpha\beta}({\bf r}) = \int \frac{{\rm d}^2k}{(2 \pi)^2} \,
G_{\alpha\beta}[{\bf k}] \exp (i {\bf k}\cdot{\bf r}).
\label{eqn:Greal}
\end{equation}
Equation (\ref{vfield}) can be rewritten as 
\begin{align}
v_\alpha ({\bf r}) &= 
\int \frac{{\rm d}^2k}{(2 \pi)^2} 
\int {\rm d} {\bf r}' \, 
\exp (i {\bf k}\cdot{\bf r}) 
G_{\alpha\beta}[{\bf k}] 
\exp (-i {\bf k}\cdot{\bf r'})
\nonumber\\& \quad\quad \quad\quad \quad\quad
\times
F_\beta^{(\ell)}({\bf r}') \frac{\delta (r'-R)}{2 \pi R}.  
\label{eq:vfields}
\end{align}
Let $\varphi$ denote the angle between ${\bf r}'$ and ${\bf U}$. 
Then we obtain 
\begin{align}
 \int {\rm d} {\bf r}'& \,  
\exp (-i {\bf k}\cdot{\bf r'})
F_\beta^{(\ell)} ({\bf r}')\frac{\delta (r'-R)}{2 \pi R} \nonumber \\
= & \int_0^{2\pi} \frac{{\rm d} \varphi}{2 \pi} \,  
\exp \left[-i \left(k_x \cos \varphi + k_y \sin \varphi \right)R \right]
F^{(\ell {\rm n})} \cos \varphi \, \frac{r'_\beta}{r'}, 
\label{reexpress}
\end{align}
where ${\bf r}'/r'$ is the unit orientational vector. 
Equation (\ref{eq:vfields}) can be rewritten by using Eq.~(\ref{reexpress}) 
and the relation     
\begin{align}
\left( \delta_{\alpha\beta} - \frac{k_\alpha k_\beta}{k^2} \right) 
\frac{r'_\beta}{r'}
= \frac{r'_\alpha}{r'}- \frac{k_\alpha}{k^2} 
\left( k_x \cos \varphi + k_y \sin \varphi \right).  
\end{align}
The integration with respect to $\varphi$ can be performed (see Appendix
for the useful relations to perform this integration), and the result 
becomes 
\begin{align}
{\bf v} ({\bf r}) &= 
\int \frac{{\rm d}^2k}{(2 \pi)^2} \,
\frac{\exp (i {\bf k}\cdot{\bf r})}{\eta [ k^2 + \nu k \coth(kh)]} 
\left[ \frac{k_y^2}{k^3 R}J_1 (kR)  \hat{{\bf e}}_x  \right.
\nonumber\\&\quad
\left. +
\left( \frac{k_y}{k^2} J_2 (kR) - \frac{k_x k_y}{k^3 R} J_1 (kR)
\right) \hat{{\bf e}}_y \right] F^{(\ell {\rm n})}.  
\label{v_1}
\end{align}

Finally, we note ${\bf k}=(k \cos \phi, k \sin \phi)$ and  
${\bf r}=(r \cos \theta, r \sin \theta)$ as well as the relation
\begin{align}
\int \frac{{\rm d}^2k}{(2 \pi)^2} \,&
\exp (i {\bf k}\cdot{\bf r}) 
= \int_0^{2\pi} \frac{{\rm d} \phi}{2\pi} 
\int_0^\infty \frac{{\rm d} k}{2\pi} \, 
\nonumber\\&\times
k \exp \left[-i k r \left(\cos\theta \cos \phi + \sin\theta \sin\phi
\right) \right].
\end{align}
Then the integration with respect to $\phi$ can be performed to obtain 
\begin{equation}
{\bf v}({\bf r}) = \hat{{\bf e}}_x 
\int_0^{\infty} {\rm d} k \, \frac{J_1 (kR) J_1 (kr)}
{2 \pi\eta k R r [ k^2 + \nu k \coth(kh)]} F^{(\ell {\rm n})}. 
\label{v_2}
\end{equation}
By identifying the velocity at the periphery of the domain to be 
${\bf U}$ and using Eq.~(\ref{eq:Gauss}), we obtain  
\begin{equation}
{\bf U} = - {\bf F} \int_0^\infty {\rm d}k \,   
\frac{J_1(kR)^2}{\pi \eta k R^2 [ k^2 + \nu k \coth(kh)]}.  
\label{v_3}
\end{equation}
The friction coefficient is given by $\zeta = -{\bf F}/{\bf U}$. 
Following the Einstein relation $D=k_{\rm B}T /\zeta$, we obtain the 
diffusion coefficient of a domain as 
\begin{equation}
D_{\rm dom} = k_{\rm B} T \int_0^\infty {\rm d}k  
\frac{J_1(kR)^2}{\pi \eta k R^2 [ k^2 + \nu k \coth(kh)]}.  
\label{eq:D}
\end{equation}
This is the generalization of the result obtained by De Koker 
to the case of finite solvent depth~\cite{DeKoker}.

\subsection{Limit of infinite thickness of the solvent layer}
\label{SaffmanDelbruck}

In this limit, the diffusion coefficient of the circular object was 
first calculated for the solid circular disk by Saffman and 
Delbr\"uck~\cite{saffman-75,saffman-76}. 
For a circular liquid domain which has the same viscosity as the 
outside of the domain, the diffusion coefficient was obtained by 
De Koker~\cite{DeKoker}.
The similar expression was obtained for the domain shape relaxation times~\cite{Camley}.

By taking the limit of $kh\gg1$ in Eq.~(\ref{eqn:cothoseen}), the 
mobility tensor can be written as~\cite{inaura-08,oppenheimer-09}
\begin{equation}
G_{\alpha\beta}[{\bf k}] =  
\frac{1}{\eta ( k^2 + \nu k)} 
\left( \delta_{\alpha\beta} - \frac{k_\alpha k_\beta}{k^2} \right).
\label{eqn:sdoseen}
\end{equation}
If the above mobility tensor is used in Eq.~(\ref{eq:D}), it reduces 
to that derived by De Koker \cite{DeKoker}. 
In this case, the integration can be performed by using Mathematica 
with the use of Meijer $G$-functions~\cite{MATHEMATICA}  
\begin{align}
D_{\rm dom}& = \frac{k_{\rm B} T }{\displaystyle 2 \pi  \eta (\nu R)^2}
\left[ -\frac{2}{(\nu R)^2} - 1\right.
\nonumber\\&\quad
\left.
- \frac{1}{\pi^{3/2} }
G_{2 \, 4}^{3 \, 2} 
\left[
(\nu R)^2 \left| 
\begin{array}{c}
1/2, 3/2 \\
0, 1, 3/2, -1
\end{array}
\right. 
\right] 
\right]. 
\end{align} 
This expression is useful to take the limits with respect to $\nu R$.

In the case of $\nu R \ll 1$, the above expression reduces to 
\begin{equation}
D_{\rm dom} \approx \frac{k_{\rm B}T}{4 \pi \eta}
\left[\ln\left(\frac{2}{\nu R}\right)-\gamma +\frac{1}{4} \right]. 
\label{eq:dES}
\end{equation}
The difference from the result by Saffman and Delbr\"uck is the
additional factor $1/4$ in the r.h.s.\ of 
Eq.~(\ref{eq:dES}) ~\cite{saffman-75,saffman-76}. 
This means that the diffusion coefficient of a circular domain is
slightly larger than that of the disk, since the flow induced inside 
the domain reduces the friction between the membrane flow and the 
domain periphery compared to that between the membrane flow and the 
solid edge. 
In the opposite limit of $\nu R \gg 1$, the diffusion coefficient is obtained as 
\begin{equation}
D_{\rm dom} \approx  \frac{4 k_{\rm B}T}{3\pi^2 \eta_{\rm s} R}, 
\label{eq:dLES}
\end{equation}
which is inversely proportional to the domain radius, $R$. 
The obtained diffusion coefficient is again slightly larger than that  
of the disk in the same limit~\cite{hughes-81,Petrov-08}
\begin{equation}
D_{\rm disk} \approx \frac{k_{\rm B}T}{8 \eta_{\rm s} R}.
\label{eq:Petrov}
\end{equation}
The fact that $D_{\rm dom}$ is inversely proportional to $R$ is consistent 
with the result of 2D polymer chain in the
membrane~\cite{Ramachandran-20}.

\subsection{The limit of thin solvent layer}
\label{EvansSackmann}

The diffusion in supported membranes in the $\nu h\ll 1$ limit was 
originally considered by Evans and Sackmann for the solid disk
immersed in the membrane~\cite{evans-88}.
The diffusion coefficient of a circular viscous domain embedded in the 
membrane was recently studied by us~\cite{sanoop-drag-10}. 
In this case, Eq.~(\ref{eqn:cothoseen}) takes the following form
\begin{equation}
G_{\alpha\beta}[{\bf k}] =  
\frac{1}{\eta ( k^2 + \kappa^2)} 
\left( \delta_{\alpha\beta} - \frac{k_\alpha k_\beta}{k^2} \right), 
\label{eqn:esoseen}
\end{equation}
where $\kappa \equiv(\nu/h)^{1/2}$.
The above mobility tensor was previously used by 
us~\cite{komura-95,seki-93,seki-07,sanoop-drag-10,Tserkovbyak-06}.

We replace $\nu k \coth(kh) \simeq \kappa^2$ for $h\rightarrow 0$ in 
the integrand of Eq.~(\ref{eq:D}).
A rigorous condition of small $h$ needs some care since 
$\kappa =(\nu/h)^{1/2}$ diverges in the limit of $h\rightarrow 0$ when 
$\nu$ is finite. 
In the previous section, we have discussed the condition in detail 
and shown that 
the results are valid under the condition given by $\nu h <1$. 
With this replacement, we obtain 
\begin{equation}
D_{\rm dom}= \frac{k_{\rm B}T}{\pi \eta (\kappa R)^2} 
\left[\frac{1}{2}-I_1(\kappa R) K_1(\kappa R) \right],
\label{eq:dE}
\end{equation}
which coincides with our previous result~\cite{sanoop-drag-10}. 
However, it should be noted that the diffusion coefficient was obtained 
by taking into account the hydrodynamic force from the membrane alone in  
Ref.~\cite{sanoop-drag-10}. 
In order to compare the present result with our previous result, the
direct friction between the solvent and the domain, $\pi \eta (\kappa R)^2$, 
should be added to the previous result. 
This leads to add $k_{\rm B}T/\pi \eta (\kappa R)^2$ to the diffusion 
coefficient.
For comparison, we also write the result by Evans and Sackmann~\cite{evans-88} 
\begin{equation}
D_{\rm ES}= \frac{k_{\rm B}T}{\pi \eta (\kappa R)^2}
\left[2+ \frac{4 K_1(\kappa R)}{\kappa R K_0(\kappa R)} \right]^{-1}, 
\label{eq:dEe}
\end{equation}
where the direct friction between the solvent and the domain is added.
As pointed out before, Eq.~(\ref{eq:dE}) is slightly larger than 
Eq.~(\ref{eq:dEe})~\cite{sanoop-drag-10}.
This is because the fluid flow in the domain reduces  
the friction between the domain and the host membrane at the edge.

In the limit of $\kappa R \ll 1$, the previous result is 
reproduced~\cite{sanoop-drag-10}
\begin{equation}
D_{\rm dom} \approx \frac{k_{\rm B}T}{4 \pi \eta}\left[ 
\ln\left(\frac{2}{\kappa R}\right)-\gamma +\frac{1}{4} \right]. 
\label{eq:dE_1}
\end{equation}
In the opposite limit of $\kappa R \gg 1$, the diffusion coefficient is 
obtained as 
\begin{equation}
D_{\rm dom} \approx \frac{k_{\rm B}T}{2 \pi \eta (\kappa R)^2}.  
\label{eq:dE_2}
\end{equation}
In this limit, $D_{\rm dom}$ decays as $1/R^2$ as pointed out 
before~\cite{sanoop-drag-10}.

\subsection{Finite thickness of solvent layer}
\label{finitedepth}

In the case of finite $h$, the integration of Eq.~(\ref{eq:D}) can be 
transformed into the summations as employed before
\begin{equation}
D_{\rm dom} =\sum_{j=1}^\infty C_j  \frac{k_{\rm B}T}{\pi \eta (\kappa_j R)^2} 
\left[ \frac{1}{2}-I_1(\kappa_j R) K_1(\kappa_j R) \right],
\label{eq:dE_main}
\end{equation}
where $C_j$ is the weight factor given by Eq.~(\ref{eq:Cj}) and 
$\kappa_j$ is determined by Eq.~(\ref{eq:kappaj}). 

When $\nu h <1$, 
$C_j$ decreases rapidly as $j$ increases as already shown in the previous section. 
In this case, Eq. (\ref{eq:dE_main}) can be approximated by the lowest order expression, 
\begin{equation}
D_{\rm dom}\approx C_1 \frac{k_{\rm B}T }{\pi \eta (\kappa_1 R)^2}
\left[ \frac{1}{2}-I_1(\kappa_1 R) K_1(\kappa_1 R) \right], 
\label{eq:dEes1}
\end{equation}
where $\kappa_1$ is the smallest positive value of $\kappa_j$.
For $\kappa_1 R \ll 1$, Eq. (\ref{eq:dEes1}) reduces to 
\begin{equation}
D_{\rm dom} \approx C_1\frac{k_{\rm B}T }{4 \pi \eta}
\left[ \ln\left(\frac{2}{\kappa_1 R}\right)-\gamma +\frac{1}{4}\right], 
\label{eq:dE_1s1}
\end{equation}
whereas for $\kappa_1 R  \gg 1$, it reduces to 
\begin{equation}
D_{\rm dom} \approx C_1 \frac{k_{\rm B}T }{2 \pi \eta (\kappa_1 R)^2}. 
\label{eq:dE_1s2}
\end{equation}

When $\nu h >1$, 
Eq. (\ref{eq:dEes1}) approximates the exact expression, Eq. (\ref{eq:dE_main}), 
only when the expression multiplied to  
$C_j$ rapidly decreases with increasing $j$ as we have already stated 
in the previous section. 
Since the hydrodynamic screening lengths and the weights factors are 
common, the difference between the polymer and the circular liquid 
domain originates from the non-dimensional size dependence. 
However, the size dependence is very similar between the polymer and 
the circular domain, i.e., weak logarithmic dependence for 
relatively small sizes and the algebraic dependence at large sizes. 
As a consequence, essentially the same results as those shown in 
Fig.~\ref{fig:if_polymer} are obtained for the two cases. 
The condition that the diffusion coefficient can be approximated by 
the expression with a single hydrodynamic screening length is
essentially the same for the polymer and the circular liquid domain. 
When $\nu h >1$, Eq. (\ref{eq:dE_1s2}) is a good approximate expression 
of Eq. (\ref{eq:dE_main}) but 
Eq. (\ref{eq:dE_1s1}) is not. 

\begin{figure}%
\centerline{
\includegraphics[width=0.5\columnwidth]{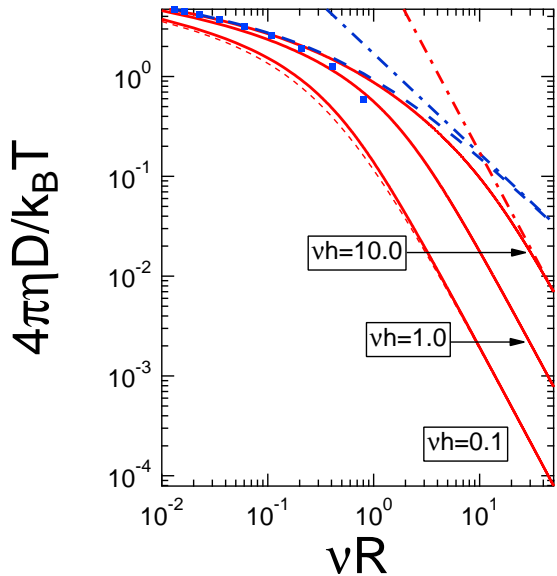}
}
\caption{ (Color online) The diffusion coefficients of a liquid domain against size 
for various solvent thickness. 
The solid red lines represent the generalized solution of De Koker 
given by Eq.~(\ref{eq:D}).  $\nu h=10$, $\nu h=1$, and $\nu h=0.1$ 
from right to left. 
The result in the limit of thin solvent layer, Eq.~(\ref{eq:dE}), 
overlaps with that of $\nu h  =0.1$. 
The red dashed-dotted line represents the asymptotic solution for 
$\nu h =10$, Eq.~(\ref{eq:dE_1s2}). 
The red thin dashed line indicates the Evans-Sackman's expression, 
Eq.~(\ref{eq:dEe}). 
The blue long dashed line represents the original solution of 
De Koker obtained by taking $h \rightarrow \infty$ limit in Eq.~(\ref{eq:D}). 
The blue dots denote the results of Eq.~(\ref{eq:dES}) and the 
blue dashed-dotted line indicates the asymptotic results, Eq.~(\ref{eq:dLES}). 
} 
\label{fig:DeKoker}
\end{figure}

Before closing the section, 
we compare  in Fig.~\ref{fig:DeKoker} the generalized solution of 
De Koker given by Eq.~(\ref{eq:D}) with the results in the two limits; 
the original solution of De Koker obtained in the limit of $h \rightarrow \infty$ 
and the results of Eq.~(\ref{eq:dE}) obtained in the limit of $h \rightarrow 0$. 
The results of Eq.~(\ref{eq:D}) with $\nu h = 0.1$ overlap with the 
results of Eq.~(\ref{eq:dE}). 
The diffusion coefficient of Eq.~(\ref{eq:dE}) is slightly larger than 
that of the solid disk, Eq.~(\ref{eq:dEe}). 
By increasing $\nu h $ the results shift toward the original solution of De Koker. 
In the asymptotic limit, the diffusion coefficient scales with $1/R^2$  
in the generalized solution of De Koker, 
while the diffusion coefficient scales with $1/R$ in the the original 
solution of De Koker. 
In the opposite limit of $R \rightarrow 0$, all the results of Eq.~(\ref{eq:D}) 
as well as the original solution of De Koker show the logarithmic dependence on $R$ 
as represented by Eq.~(\ref{eq:dES}).

\section{Conclusions}
\label{conclusions}

The diffusion coefficient of an inclusion in a membrane is 
strongly influenced by the presence of solvents due to the stick 
boundary condition between the membrane and the solvent. 
The thickness of solvent layer is a key parameter controlling the 
diffusion of an inclusion in a membrane. 
In this work, the diffusion coefficient of a polymer confined in 
a membrane is obtained for arbitrary thickness of solvents. 
We also study the influence of finite thickness of solvent on the 
diffusion coefficient of a circular liquid domain with the same 
viscosity as that of the host membrane. 
Previously, the diffusion coefficient of a circular liquid domain was 
expressed by a single integral in the limit of 
infinite solvent thickness~\cite{DeKoker}. 
In this work, the integral expression is generalized to the case of 
finite solvent depth. 
The various analytical expressions are obtained from the integral expression.

In general, the diffusion coefficient of inclusions decreases as the 
solvent thickness decreases. 
The solvent induces drag against the membrane flow, and as a result 
diffusion is suppressed. 
The drag increases as the distance between the membrane and the
substrate decreases. 
More precisely, the influence of solvent on the membrane flow is
characterized by the hydrodynamic screening lengths. 
Multiple hydrodynamic screening lengths can be obtained from the characteristic equation Eq.~(\ref{eq:kappaj}), when the solvent thickness 
is nonzero. 
The largest hydrodynamic screening length characterizes the length scale 
of momentum dissipation from the membrane to the solvent. 
The membrane flow dissipates the momentum to the solvent through the 
stick boundary condition. 
For finite solvent depth, the largest hydrodynamic screening length 
is approximately given by $\sqrt{h/\nu}$. 
The diffusion coefficients are expressed by the weighted sum of the 
functions of the non-dimensional size of the inclusion normalized 
by the hydrodynamic screening lengths. 
The weights are given by Eq.~(\ref{eq:Cj}).

By examining both the weights and the non-dimensional size dependence, 
we find that the diffusion coefficient can be approximated by the 
expression given by the largest hydrodynamic screening length alone 
except when both $\nu^{-1}$ and the size of the inclusion 
are smaller than the solvent thickness. 
 (The results are summarized in Table~\ref{tab:1}.) 
When the solvent thickness is larger than $\nu^{-1}$, $h > \nu^{-1}$, the weights 
decrease slowly with increasing $j$. 
Hence $\nu^{-1}$ can be viewed as the critical solvent depth below 
which there is a dominant mode given by the largest hydrodynamic 
screening length. 
If the size of the inclusion is smaller than the solvent thickness, 
$\kappa_1 R$ is close to $\kappa_2 R$ since we have 
$\kappa_2 R- \kappa_1 R \sim R/h$.  
Besides, if the diffusion coefficient depends weakly on $\kappa_j R$, 
the diffusion coefficient expressed by $\kappa_1 R$ alone cannot
represent the exact summed result. 
In particular, the sum of logarithmic functions of screening lengths 
weighted by the similar factors cannot be represented by one of the terms. 
The situation is met when both $\nu^{-1}$ and the size of the inclusion 
are smaller than the solvent thickness.

The results summarized above are obtained for the finite solvent thickness. 
In the limit of $h \rightarrow \infty$, a new length scale appears as 
discussed by Diamant~\cite{diamant-09b}. 
The diffusion coefficient is given by the new hydrodynamic screening 
length $\nu^{-1}$.

\begin{table}
\caption{Summary of the size dependence of the diffusion coefficient when the solvent depth is finite.}
\label{tab:1}      
\begin{tabular}{c|c|c|c|c}
\hline\noalign{\smallskip}
amplitude separation & screening lengths & solvent depth & domain size & 
size dependence \\
($h<1/\nu$) &separation ($h<R$)&  &  & 
 \\
\hline
complete & \textemdash\footnotemark[1] & $h<1/\kappa_1<1/\nu$ &$R<1/\kappa_1$&
$\ln(\kappa_1 R)$ \\
complete  & \textemdash\footnotemark[1]   &$h<1/\kappa_1<1/\nu$&$1/\kappa_1<R$
& $1/(\kappa_1 R)^2$ \\
incomplete  & incomplete & $1/\nu<1/\kappa_1<h$&$R<h$&
 \textemdash\footnotemark[2]   \\
incomplete & complete  &$1/\nu<1/\kappa_1<h$& $h<R$
 & $1/(\kappa_1 R)^2$ \\
\noalign{\smallskip}\hline
\end{tabular}
\footnotetext[1]{
The summation in the expression of the diffusion coefficient can be approximated by the dominant term as long as amplitude separation is complete 
no matter about the separation of screening lengths. 
} 
\footnotetext[2]{The summation in the expression of the diffusion coefficient cannot be approximated by a single dominant term. 
Even in the regime of $1/\kappa_1<R<h$, $1/(\kappa_1 R)^2$ spatial dependence is not obtained. } 
\end{table}

For relatively small inclusions, the diffusion coefficients can be 
approximated by the logarithmic function of the size normalized by 
ES hydrodynamic screening length $\sqrt{h/\nu}$ when 
the solvent depth is small, i.e, $1/\nu >h$. 
In the opposite limit of $h \rightarrow \infty$, the diffusion 
coefficient is given by the logarithmic function of the size 
normalized by $\nu^{-1}$. 
In the intermediate solvent depth, the diffusion coefficients are 
expressed by the sum of multiple terms and the diffusion coefficient 
cannot be represented by the logarithmic function of the largest 
hydrodynamic screening length.  
However, the diffusion coefficient may be approximated by a logarithmic 
function.
This directs us to define an empirical interpolation of 
the effective hydrodynamic screening lengths 
by the inverse of $\kappa^*= \nu/(1+ \sqrt{\nu h})$ for any value 
of $h$ when inclusions are small.

The size dependence of diffusion coefficients is influenced by the 
solvent depth. 
In the case of a supported membrane, the typical value of $h$ is 
$20$ \AA \mbox{} and $\nu h \sim 10^{-2}$ can be estimated by
introducing typical values of membranes;  
$\eta_{\rm s}=10^{-2}$ poise and $\eta$ given by 
$1$ poise multiplied by the membrane thickness 
$5 \times 10^{-3}$ $\mu$m~\cite{Tanaka-07}. 
This is the case when the ES hydrodynamic screening length 
Eq.~(\ref{eq:dEe}) or its modification Eq.~(\ref{eq:dE}) are relevant. 
By using relatively large size of inclusion, $R> h=20$ \AA\mbox{ }, 
the asymptotic $1/R^2$ dependence of diffusion coefficient can be
observed.

In the case of vesicles of $10$ $\mu$m size, $\nu h \simeq 1$ can be 
estimated by interpreting vesicle radius as solvent 
thickness~\cite{yanagisawa-07}. 
Since the inclusion is normally smaller than the vesicle radius, 
we have $R/h <1$. 
In this case, we can estimate as 
$\kappa R  \sim \nu R/\sqrt{\nu h} < \sqrt{\nu h} \sim 1$.  
If the vesicle radius is regarded as a solvent thickness, 
it may be difficult to observe the 
asymptotic $1/R^2$ dependence of the diffusion coefficient. 
However, this estimation is not rigorous but 
is done just for the purpose of indicating the boundary effect caused by the finite radius of a vesicle. 
The real flow inside a vesicle should be different from that in the presence of the solid substrate. 
It should be also reminded that there is an additional difficulty to 
differentiate the translational diffusion of a domain from 
the rigid rotation of the vesicle~\cite{Aliaskarisohi}.

For simplicity, we have considered the situation where the membrane is 
floated on a solvent layer. 
In general, both sides of a membrane are surrounded by solvents. 
We consider the case that the solvent layer on the membrane is covered 
by a substrate and is not a free standing film. 
We denote the 3D viscosity of solvent and the thickness in the upper 
domain as $\eta_{\rm s}^+$ and $h^+$, respectively, and 
those in the lower domain as $\eta_{\rm s}^-$ and $h^-$, respectively.   

Equation (\ref{eqn:cothoseen}) and Eq. (\ref{eq:D}) are valid if we replace 
$k + \nu  \coth (kh)$ with 
$k + \nu^+  \coth (kh^+)+ \nu^-  \coth (kh^-)$, 
where $\nu^+=\eta_{\rm s}^+/\eta$ and 
$\nu^-=\eta_{\rm s}^-/\eta$ ~\cite{sanoop-11}.  
Correspondingly, the characteristic equation becomes 
\begin{equation}
\nu^+ \cot (\kappa_j h^+)+\nu^- \cot (\kappa_j h^-) = \kappa_j . 
\label{eq:kappajpm}
\end{equation}
The influence of solvents on both sides of the membrane can be 
investigated by studying the roots of Eq.~(\ref{eq:kappajpm}). 
By using $\cot x \approx 1/x$, we obtain 
\begin{equation}
\kappa_1= \sqrt{
\frac{\nu^+}{h^+}+\frac{\nu^-}{h^-}} ,  
\label{eq:kappa1pm}
\end{equation}
when $\kappa_1 h^+<1$ and $\kappa_1 h^- <1$.

In the simple situation where the membrane is sandwiched by the same 
depth of solvent layers $h=h^+=h^-$, both the characteristic equation, 
Eq.~(\ref{eq:kappaj}) and the weights given by Eq.~(\ref{eq:Cj}) still 
hold by using  the renewed definition of $\nu =\nu^+ + \nu^-$. 
The diffusion coefficients in this particular case can be obtained 
from those presented in this manuscript by substituting 
$\nu = \nu^+ + \nu^-$ ~\cite{sanoop-drag-10}.

If both $h^+$ and  $h^-$ are infinite, the screening length changes 
from $\kappa_1^{-1}$ to $(\nu^+ + \nu^-)^{-1}$. 
Our study on the influence of the finite solvent thickness indicates 
that the diffusion coefficient can be approximately expressed in terms 
of  $\kappa_1$ except when both the SD screening length and 
the size of the inclusion are smaller than both $h^+$ and $h^-$.

\acknowledgements

We would like to thank Youhei Fujitani for fruitful discussions. 
This work was supported by KAKENHI (Grant-in-Aid for Scientific
Research) on Priority Area ``Soft Matter Physics'' and Grant
No.\ 21540420 from the Ministry of Education, Culture, Sports, 
Science and Technology of Japan.

\renewcommand{\theequation}{A.\arabic{equation}}  
\setcounter{equation}{0}  
\section*{Appendix: Useful relations}

We have used the relations, 
\begin{equation}
\int_0^{2 \pi}  \frac{{\rm d} \varphi}{2\pi} \,
\exp \left[-i \left(k_x \cos \varphi + 
k_y \sin \varphi \right)a \right]= J_0 (ka)
\end{equation}
and 
\begin{align}
\int_0^{2 \pi}&  \frac{{\rm d} \varphi}{2\pi} \,
\exp \left[-i \left(k_x \cos \varphi + 
k_y \sin \varphi \right)a \right] \cos^2 \varphi &
\nonumber\\ &
=- \frac{1}{a^2} \frac{\partial^2}{\partial k_x^2} J_0 (ka) 
= - \frac{k_x^2}{k^2} J_2 (ka) + \frac{J_1 (ka)}{ka},
\end{align}
\begin{align}
\int_0^{2 \pi} & \frac{{\rm d} \varphi}{2\pi} \, 
\exp \left[-i \left(k_x \cos \varphi + 
k_y \sin \varphi \right)a \right] \cos \varphi \sin \varphi &
\nonumber\\ &
= - \frac{1}{a^2} \frac{\partial^2}{\partial k_x \partial k_y} J_0 (ka) 
=- \frac{k_x k_y}{k^2} J_2 (ka). 
\end{align}



\end{document}